\documentclass[12pt]{article}
\usepackage{bm}
\usepackage{indentfirst}
\usepackage{natbib}
\usepackage{graphics}
\usepackage{graphicx}
\usepackage{amsmath}
\usepackage{verbatim}
\usepackage{multirow}
\usepackage{float}
\usepackage{tikz}
\usepackage{amsmath, nccmath}
\usepackage[margin=1in]{geometry}
\usepackage[colorlinks=true,linkcolor=blue,urlcolor=black,bookmarksopen=true,citecolor=blue]{hyperref}
\usepackage{bookmark}
\usetikzlibrary{decorations.pathreplacing}
\usepackage{setspace}
\usepackage{comment}
\usepackage{adjustbox}
\usepackage{xcolor}
\usepackage{chngcntr}
\usepackage{authblk}
\usepackage{algorithm}
\usepackage{algpseudocode}
\usepackage{caption}
\usepackage{textcomp}

\newcounter{algsubstate}
\renewcommand{\thealgsubstate}{\alph{algsubstate}}
\newenvironment{algsubstates}
{\setcounter{algsubstate}{0}%
	\renewcommand{\State}{%
		\stepcounter{algsubstate}%
		\Statex {\footnotesize\thealgsubstate:}\space}}
{}

\begin{document}
\title{A Bayesian Hidden Semi-Markov Model with Covariate-Dependent State Duration Parameters for High-Frequency Environmental Data}
\date{}
\author[1]{Shirley Rojas-Salazar}
\author[1]{Erin M. Schliep}
\author[1]{Christopher K. Wikle}
\author[2]{Emily H. Stanley}
\author[2]{Stephen R. Carpenter}
\author[2]{Noah R. Lottig}
\affil[1]{Department of Statistics, University of Missouri}
\affil[2]{Center for Limnology, University of Wisconsin}

\maketitle

\doublespacing

\begin{abstract}
Environmental time series data observed at high frequencies can be studied with approaches such as hidden Markov and semi-Markov models (HMM and HSMM). 
HSMMs extend the HMM by explicitly modeling the time spent in each state. In a discrete-time HSMM, the duration in each state can be modeled with a zero-truncated Poisson distribution, where the duration parameter may be state-specific but constant in time. We extend the HSMM by allowing the state-specific duration parameters to vary in time and model them as a function of known covariates observed over a period of time leading up to a state transition. 
In addition, we propose a data subsampling approach given that high-frequency data can violate the conditional independence assumption of the HSMM. 
We apply the model to high-frequency data collected by an instrumented buoy in Lake Mendota. We model the phycocyanin concentration, which is used in aquatic systems to estimate the relative abundance of blue-green algae, and identify important time-varying effects associated with the duration in each state.
\end{abstract}
Keywords: HSMM; MCMC; data subsampling; cyanobacteria, phycocyanin; Lake Mendota

\section{Introduction}
\label{sec:int}

Environmental time series data are measured at different frequencies but have been increasingly obtained at high temporal resolutions. These high-frequency data can be studied using a wide range of analyses. For example, \cite{li_multi-site_2021} presented a stochastic precipitation generator and applied it to high-frequency (30-second) rainfall data. \cite{lin_wind_2020} used high-frequency (1-second) supervisory control and data acquisition (SCADA) wind power data to predict power. They utilized a deep neural network to predict the wind power and incorporated the isolation forest method to identify anomalies in the data points.

High-frequency time series data in lakes provide another interesting example, and have been analyzed with different models and statistical approaches.
\cite{carpenter_stochastic_2020} studied the dynamics of cyanobacteria in Lake Mendota using a drift-diffusion-jump model. The model was applied to phycocyanin concentrations measured every minute for the years 2008 through 2018. They found that for each of the years studied, the concentration of phycocyanin can be summarized with two stable states. Another example is the study by \cite{coloso_difficulty_2011}, who looked at the drivers of lake ecosystem metabolism. They fitted a multiple linear regression to high-frequency data from two temperate lakes. For each of the dependent variables, gross primary production, respiration, and net ecosystem production, different important drivers were identified, including temperature, wind speed, photosynthetically active radiation, among others.

Hidden Markov and hidden semi-Markov models provide an alternative approach for analyzing high-frequency environmental data. A hidden Markov model (HMM) consists of a sequence of unobserved discrete states and another set of observable random variables that are assumed conditionally independent given the state at each observed time point \citep{rabiner_tutorial_1989}. The transition from one state to another depends on a transition probability, which is defined conditionally on the current state, and where the probability of self-transitioning (i.e., remaining in the same state) is non-zero. This non-zero probability implies that the time spent in each state follows a geometric distribution \citep{yu_hidden_2010}. However, this distributional assumption may not be realistic for some processes, making it necessary to additionally model the state duration. This model extension defines the hidden semi-Markov model (HSMM) \citep{yu_hidden_2016}.

Environmental data have been modeled with HMMs and HSMMs. For example, \cite{rousseeuw_hybrid_2015} applied a hybrid HMM to model phytoplankton dynamics using data measured every 20 minutes in a marine station. They incorporated a spectral clustering method into their HMM modeling in order to build a fully unsupervised HMM. 
\cite{stoner_advanced_2020} developed an HMM to analyze sub-daily rainfall data, and, through the use of simulations, were able to show that their model can capture characteristics such as long dry periods or seasonal variation. They also applied the model to a real dataset of hourly time series of rainfall in Exeter, UK.
Similar types of data have been analyzed with semi-Markov and HSMMs. For example, \cite{king_semi-markov_2016} present an extension of the Arnason-Schwarz model where they define a semi-Markov model for the state process, and apply the model to capture-recapture data of house finches.
\cite{sansom_spatial_2008} studied the spatial and temporal variation of rainfall with an HSMM using a high temporal resolution rainfall dataset collected in New Zealand.

When the duration in an HSMM is modeled with a Poisson distribution, the duration parameter, which can be different for each hidden state, is assumed to be constant in time. This assumption, however, might not be reasonable in all cases. If we consider, for example, hourly rainfall data observed over the course of a year, and we model it with two different states representing wet and dry episodes, we would expect the length of time of these episodes to be different depending on the time of year due to seasonal rainfall patterns (e.g., monsoon season). 

To capture this temporal variation in the duration in each state, we extend the HSMM by modeling the duration parameters as a function of time varying covariates.
This enables the identification of factors associated with the time spent in the different states. For example, when there is a state transition, the duration parameter for the new state could be modeled as a function of covariates observed in the period leading up to the transition, or the value of the covariate at the moment right before the switch. The functional relationship between covariates and the parameters of the duration distribution could be state-specific, and modeling these relationships can provide important inference with regard to their extent and direction.
Importantly, inference is not obtained at the high-frequency level at which the data are collected, rather it is obtained in terms of the duration intervals.

In both HMMs and HSMMs, observations are described with an emission distribution and are assumed to be conditionally independent given the state, meaning they are independent of previous states and observations \cite[]{pohle_selecting_2017,yu_hidden_2016}. However, high-frequency data are more likely to be correlated. The violation of the conditional independence assumption can have dramatic impacts on statistical inference.  \cite{pohle_selecting_2017} presented a simulation study to determine the effects of assumption violations in the selection of the number of states in HMMs, and determined that when the conditional independence assumption is violated, and the Akaike and Bayesian information criteria are used to select between models, the number of states will be overestimated. 

Markov-switching regression models (MSR) and neural networks (NN) are two approaches that have been used when the conditional independence assumption is not met. In MSR, the observations are modeled as a function of covariates or as an autoregressive model \citep{langrock_markov-switching_2017}. In the context of NN, \cite{ravuri_how_2016} apply a deep neural network HMM (DNN-HMM) and show that deeper NNs compensate for the conditional independence assumption violation more than shallow NNs. In addition, \cite{dai_recurrent_2017} consider a recurrent hidden semi-Markov model (R-HSMM) that incorporates a recurrent neural network (RNN) in the observation model of an HSMM to accommodate more complex dependencies in the observation sequence.

The disadvantage of approaches such as NN is that they are computationally expensive to implement. Thus, it is useful to consider a more computationally tractable approach for mitigating the conditional dependence. One approach that has not been considered in this context is data subsampling. Data subsampling is used for reducing computational cost or in determining the sampling distribution of a statistic. For instance, it has been used as an alternative to deal with large datasets to increase efficiency in Firefly Monte Carlo (FlyMC) \citep{maclaurin_firefly_2015} and in subsampling Markov chain Monte Carlo (MCMC) \citep{quiroz_speeding_2019}. Both of these approaches present an MCMC sampling algorithm that considers only a subset of the data at each iteration. Experiments conducted to evaluate the performance of FlyMC found it to be more efficient than standard MCMC sampling in many instances. Similarly, \cite{quiroz_speeding_2019} showed that subsampling MCMC is often more efficient than standard MCMC and other competing subsampling algorithms. These studies suggest that, for an HSMM, random data subsampling can be introduced as part of the MCMC algorithm as an attempt to reduce or eliminate the conditional dependence in the data.

The goal of this paper is to develop an HSMM with time-varying duration parameters that are dependent on covariates for studying high-frequency environmental data. Specifically, we model high-frequency concentrations of phycocyanin, an estimate of presence or relative abundance of cyanobacteria (blue-green algae), in Lake Mendota, Wisconsin. Understanding the temporal variation in cyanobacteria concentration in an urban lake is important to public health, since concentrations of blue-green algae can produce toxins that are linked to illness in humans and animals. We use covariates to explain the variation in the duration in each state and obtain inference on important characteristics. Previous approaches using HMMs or HSMMs have included covariates in the observation model or in the specification of transition probabilities  \cite[e.g.][]{koki_forecasting_2020,economou_mcmc_2014,titman_semi-markov_2010}, but the inclusion of covariates in the model for state durations has not been considered. Additionally, we propose a data subsampling approach to mitigate the violation of the conditional independence assumption that is common in high frequency data.

In Section \ref{sec:data} we present the phycocyanin data used in this analysis. Section \ref{sec:model} provides a description of the HSMM, which defines the duration distribution parameter as a function of covariates, as well as the details of a simulation study to investigate the effects of dependence in the observation sequence. The results of the model applied to the phycocyanin dataset are presented in Section \ref{sec:appl}. Lastly, Section \ref{sec:disc} provides a discussion and conclusion, as well as future extensions.

\section{High-frequency Lake Mendota data}
\label{sec:data}

The high-frequency environmental data in this application correspond to measurements from Lake Mendota, Wisconsin, in 2018. The data can be found in the North Temperate Lakes Long-Term Ecological Research program database \citep{lead_pi_north_2020}. Sensors in an instrumented buoy located in the lake recorded measurements every minute. In 2018, the buoy recorded observations from April 11 to November 15. The dataset consists of several other variables including weather conditions (air temperature, relative humidity, wind speed, and wind direction), and lake characteristics (such as chlorophyll, photosyntetically-active radiation, dissolved oxygen, etc).   

Phycocyanin is a pigment of cyanobacteria, and provides an estimate or diagnostic of its presence and concentration \citep{carpenter_stochastic_2020}. Given that high concentrations of cyanobacteria are a major public health concern, particularly in urban lakes such as Lake Mendota, it is essential to understand the temporal variation in concentration levels as well as possible environmental drivers of this variation. For each year of their study, \cite{carpenter_stochastic_2020} modeled the phycocyanin concentrations and identified two regimes of low and high phycocyanin, as well as abrupt transitions between the states. The objective of our analysis is to extend on their work by describing the latent states of cyanobacteria as captured by phycocyanin concentration, the duration in each of those states, and the covariates associated with that duration.

\begin{figure}[H]
	\begin{center}
		\includegraphics[scale=.55]{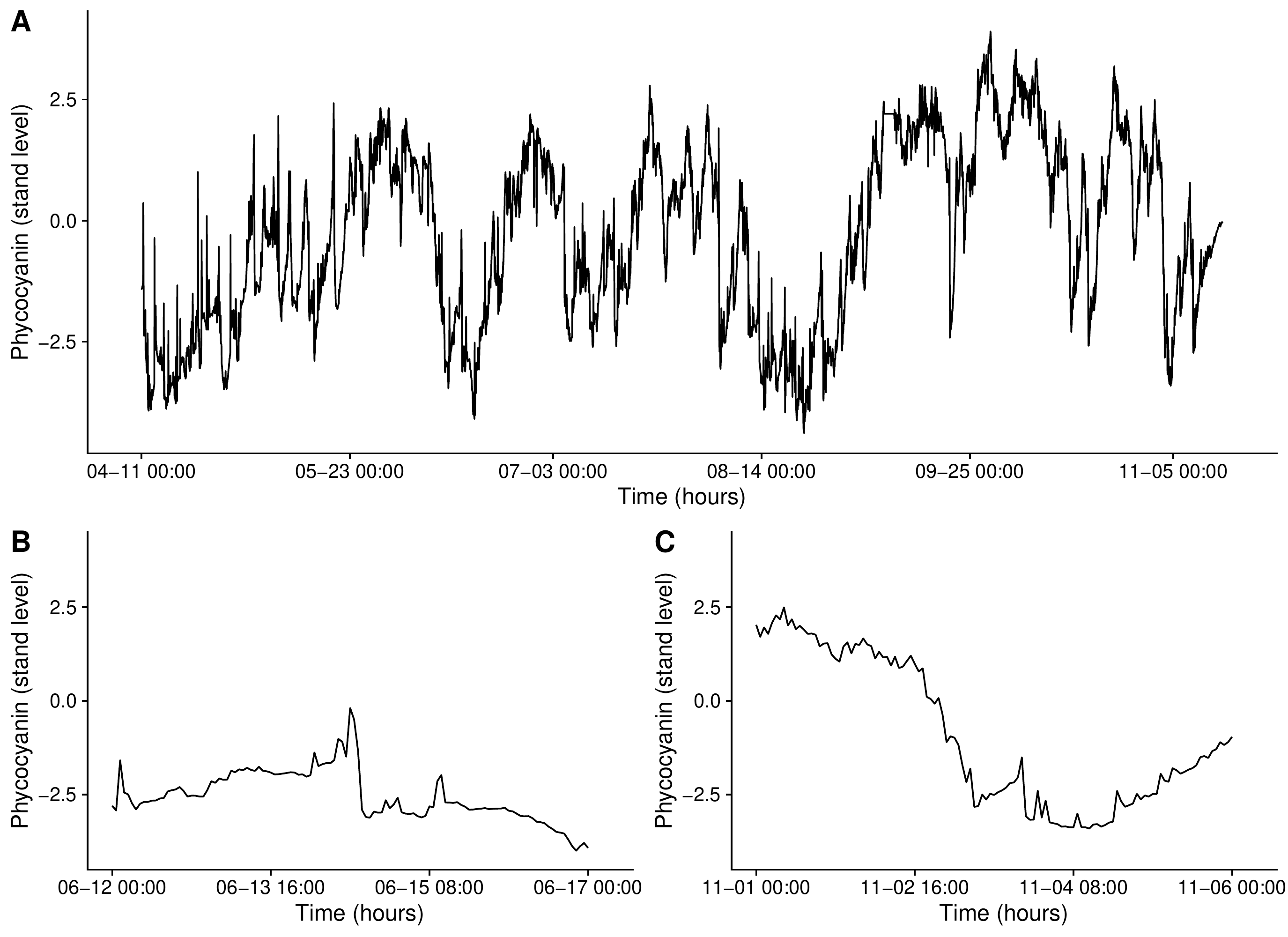}
		\caption{Phycocyanin standardized levels in Lake Mendota, 2018. Panel \textbf{A}. Full period (mid April to mid November). Panels \textbf{B} and \textbf{C} show 5-day periods in June and November, respectively. \label{FigureG3}}
	\end{center}
\end{figure}

Following the methods in \cite{carpenter_stochastic_2020}, we consider the standardized levels of phycocyanin as our observation sequence. We first compute the maximum hourly measurement, which results in a total of 5232 observations (Figure \ref{FigureG3}). These maximum values of phycocyanin are measured in relative fluorescent units (RFU). They are standardized by being log transformed ($log_{10}$), centered and scaled, and fitted using a dynamic linear model. See the supplementary information in \cite{carpenter_stochastic_2020} for more details. 


In a study done in Lake Mendota, \cite{soranno_factors_1997} found that weather variables can impact the dynamics of algae at finer time scales. Considering this, the variables we examine as possible covariates to capture the time variation in the state durations are hourly average air temperature, wind speed, relative humidity and photosynthetically-active radiation (PAR). During the time period April 11 to November 15, 2018, the air temperature, measured in $^{\circ}$C, ranged from -7 to 33, with a mean of 17. The average and maximum wind speed were 3.8 and 15 m/s respectively. Relative humidity ranged from 21\% to 100\% with an average of 74\%. Radiation was measured with a surface sensor in \textmu$\mathrm{ mol \hspace{0.5mm} m^{-2} \hspace{0.5mm} s^{-1}}$  and it ranged from 0 to nearly 2000.

\section{Hidden semi-Markov model with covariate-dependent duration parameters}
\label{sec:model}
We begin by specifying the HMM and important notation. Then, we extend the HMM to the HSMM, develop the state-specific duration model as a function of covariates, and describe the methods for Bayesian inference. Lastly, we propose a simulation to investigate the implications of the violation of the conditional independence assumption.

\subsection{Hidden Markov model and notation}
\label{subsec:HMM}
An HMM includes two stochastic processes, one represents a Markov chain of states that are hidden, and the other generates a sequence of observations that are influenced by the unobservable states \citep{rabiner_tutorial_1989,yu_hidden_2010}.
In a discrete-time HMM, the sequence of observations from time $1$ to $n$ can be denoted as $\mathbf{y}= (y_{1}, \dots,y_{n}) ^\prime$. The corresponding sequence of unobserved states is denoted as $\mathbf{S}= ( S_{1},\dots,S_{n}) ^\prime$, where $S_{i} \in \left\lbrace 1,2, \dots, M \right\rbrace $, $i=1,\dots,n$, and $M$ is the total number of unique states. The state at time $1$ has a distribution defined by $\rho_j=p\left[ S_{1}=j\right]$, $j=1,\dots,M$. The transition to the next state, $S_{2}$, is conditional on state $S_{1}$ according to the Markov property. In general, the transition probability matrix $\bf{P}$ provides the probabilities of transitioning from one state to another when the state space is discrete and constant in time. The matrix $\bf{P}$, has entries $p_{j,k}$, with $p_{j,k}=p\left[ S_{i+1}=k \mid S_{i}=j \right]$, where $1 \leq j,k \leq M$, and $\sum_{k=1}^{M} p_{j,k} =1 $.

\bigskip

\bigskip

Observations are emitted by each of the states in the hidden sequence (Figure \ref{Figurediagram}A) following a state-dependent probability distribution $f(\mathbf{y}|\bm{\theta},\mathbf{S})$. Assuming the observation distribution is Gaussian, the parameters $\bm{\theta}$ correspond to the mean and variance for each state: $\bm{\mu}=(\mu_1,\cdots,\mu_M)^\prime$ and $\bm{\sigma}^2=(\sigma^2_1,\cdots,\sigma^2_M)^\prime$. The joint likelihood of the observations can be written as:

\begin{equation*}
	L(\mathbf{y} \mid \bm{\mu}, \bm{\sigma}^2, \mathbf{S})=\prod_{i=1}^{n} f(y_{i}|\mu_{S_{i}},\sigma_{S_{i}}^2),
\end{equation*}

\noindent and the likelihood of the Markov chain is:
\begin{equation*}
	L(\mathbf{S} \mid \bm{\rho}, \mathbf{P})= \rho_{S_{1}}  \prod\limits_{i=1}^{n-1} p_{S_{i},S_{{i+1}}}.
\end{equation*} 

The complete likelihood of the Markov model is the joint likelihood of observations and states:
$L(\mathbf{y}, \mathbf{S} \mid \bm{\mu}, \bm{\sigma}^2, \bm{\rho}, \mathbf{P}) = L(\mathbf{y} \mid \bm{\mu}, \bm{\sigma}^2, \mathbf{S}) \times L(\mathbf{S} \mid \bm{\rho}, \mathbf{P})$. In summary, an HMM with $M$ states and $n$ observations has a set of model parameters  that includes the emission distribution parameters $\bm{\mu}$ and $\bm{\sigma}^2$, the initial distribution probabilities $\bm{\rho}$, and the transition probability matrix $\mathbf{P}$.

\bigskip

\begin{figure}[H]
	\centering
	
	\begin{tikzpicture}
		
		\node at (-1.0, 1.0) {\textbf{A}};
		
		\node at (0.0, 0.0) {\textit{States}};
		\node at (0.0, -1.0) {\textit{Observations}};
		
		\node (St1) at (1.8, 0.0) {$S_{{1}}$};
		\node (ys1) at (1.8, -1.0) {$y_{{1}}$};
		\draw[->] (St1) to (ys1);
		\node (St2) at (3.8, 0.0) {$S_{{2}}$};
		\node (ys2) at (3.8, -1.0) {$y_{{2}}$};
		\draw[->] (St2) to (ys2);
		
		\node (St3) at (5.8, 0.0) {$S_{{3}}$};
		\node (ys3) at (5.8, -1.0) {$y_{{3}}$};
		\draw[->] (St3) to (ys3);
		
		\draw[->] (St1) to (St2);
		\draw[->] (St2) to (St3);
		\node (dotss) at (6.8, 0.0) {$\cdots$};
		\node (dotsy) at (6.8, -1.0) {$\cdots$};
		
		\node (Stn) at (7.8, 0.0) {$S_{{n}}$};
		\node (ysn) at (7.8, -1.0) {$y_{{n}}$};
		\draw[->] (Stn) to (ysn);
		
		
		\node at (-1.0, -2.0) {\textbf{B}};
		
		\node at (0.0, -3.0) {\textit{States}};
		\node (S1) at (1.8, -3.0) {$S_{1}$};
		\node (S2) at (4.7, -3.0) {$S_{2}$};
		\node (S3) at (9.6, -3.0) {$S_{Q}$};
		
		\draw[->] (S1) to (S2);
		\node (dotss) at (8.4, -3.0) {$\cdots$};
		\node (dotsy) at (8.4, -4.0) {$\cdots$};
		
		\node at (0.0, -4.0) {\textit{Observations}};
		
		\node (y11) at (1.8, -4.0) {$y_{{1}}$};
		\node (y12) at (2.4, -4.0) {$y_{{2}}$};
		\node (dots11) at (2.9, -4.0) {$\cdots$};
		\node (yd1) at (3.5, -4.0) {$y_{{\tau_1}}$};
		\draw[->] (S1) to (y11);
		\draw[->] (S1) to (y12);
		\draw[->] (S1) to (yd1);
		
		\node (y21) at (4.7, -4.0) {$y_{{T_1+1}}$};
		\node (y22) at (5.8, -4.0) {$y_{{T_1+2}}$};
		\node (dots2) at (6.6, -4.0) {$\cdots$};
		\node (yd2) at (7.5, -4.0) {$y_{{T_1+\tau_2}}$};
		\draw[->] (S2) to (y21);
		\draw[->] (S2) to (y22);
		\draw[->] (S2) to (yd2);
		
		\node (y31) at (9.6, -4.0) {$y_{{T_{Q-1}+1}}$};
		\node (y32) at (11.0, -4.0) {$y_{{T_{Q-1}+2}}$};
		\node (dots3) at (12.05, -4.0) {$\cdots$};
		\node (yd3) at (13.1, -4.0) {$y_{{T_{Q-1}+\tau_Q}}$};
		\draw[->] (S3) to (y31);
		\draw[->] (S3) to (y32);
		\draw[->] (S3) to (yd3);
		
		\node at (0.0, -5.2) {\textit{Durations}};
		\draw [decorate,decoration={brace,raise=2mm,amplitude=3pt,mirror}] (y11.south west) -- (yd1.south east);
		\draw [decorate,decoration={brace,raise=2mm,amplitude=3pt,mirror}] (y21.south west) -- (yd2.south east);
		\draw [decorate,decoration={brace,raise=2mm,amplitude=3pt,mirror}] (y31.south west) -- (yd3.south east);
		\node (tau1) at (2.8, -5.2) {$\tau_1$};
		\node (tau2) at (6.2, -5.2) {$\tau_2$};
		\node (tau3) at (11.5, -5.2) {$\tau_Q$};
		
		\node (dotss) at (8.4, -5.2) {$\cdots$};

	\end{tikzpicture}
	
	\caption{State and observation sequences. Panel \textbf{A}. HMM: One observation is emitted by each state in the sequence. Panel \textbf{B}. HSMM: Several observations are emitted by each state, the number is determined by the duration in the state.} \label{Figurediagram}
\end{figure}

\subsection{Hidden semi-Markov model}
\label{subsec:HSMM}

Figure \ref{Figurediagram}B illustrates the HSMM where instead of assuming there is only one observation per state, a sequence of observations are emitted. The number of observations depends on the amount of time spent in the state. Following the notation in \cite{economou_mcmc_2014}, let $\tau$ represent the length of time that the sequence remains in a state before transitioning. These \emph{durations} are labeled in Figure \ref{Figurediagram}B as $\tau_1, \dots, \tau_Q$, where Q is the number of intervals or segments. 
For $q=1, \dots, Q$ we define $T_q$ to be the cumulative duration in segments $1$ through $q$.
Lastly, we define $h_j(\tau \mid \phi_j)$ as the duration distribution for each state $j$, $j=1,\dots,M$, with parameter $\phi_j$. 

Similar to the Markov model, the likelihood of the semi-Markov model has two main components consisting of the likelihood of the observations conditional on the states and the likelihood of the semi-Markov chain of states. The joint likelihood of the observations can be specified analogous to the HMM case, but is written incorporating the segment-specific notation: 

\begin{equation}
	L(\mathbf{y}|\bm{\mu}, \bm{\sigma}^2, \mathbf{S})= \prod_{i=1}^{n} f(y_{i}|\mu_{S_{i}},\sigma_{S_{i}}^2)= \prod_{q=1}^{Q} f(\mathbf{y}_{\tau_q}|\mu_{S_{q}},\sigma_{S_{q}}^2), 
\end{equation}

\noindent where $\mathbf{y}_{\tau_q}$ corresponds to the vector of all the observations in time interval $q$. The likelihood of the state sequence includes the distribution of the first state, the transition probabilities for the state switches, as well as the information from the duration times:

\begin{equation}
	\label{LMC}
	L(S_1,\dots,S_Q, \tau_1,\dots,\tau_Q | \bm{\rho}, \mathbf{P},  \bm{\phi})= \rho_{S_1}  \prod\limits_{q=1}^{Q-1}  h_{S_q}(\tau_q \mid \phi_{S_q})  p_{S_q,S_{q+1}}   h_{S_Q}(\tau_Q \mid \phi_{S_Q}).
\end{equation} 

\noindent Thus, the joint distribution of data, states and durations of the hidden semi-Markov model can be written as:

\bigskip

\hspace{2mm} $L(\mathbf{y}_{\tau_1},\dots,\mathbf{y}_{\tau_Q}, S_1,\dots,S_Q, \tau_1,\dots,\tau_Q \mid  \bm{\mu}, \bm{\sigma}^2,\bm\rho,\mathbf{P},\bm{\mathbf\phi})$

\begin{equation}
	\label{hsmmlik}
	\begin{gathered}
		\hspace{50mm} =\rho_{S_1}  \prod\limits_{q=1}^{Q-1}  h_{S_q}(\tau_q |\phi_{S_q})  p_{S_q,S_{q+1}}  f(\mathbf{y}_{\tau_q}|\mu_{S_{q}},\sigma_{S_{q}}^2)   \\
		\hspace{32mm}  \times \hspace{2mm} h_{S_Q}(\tau_Q |\phi_{S_Q})  f(\mathbf{y}_{\tau_Q}|\mu_{S_{Q}},\sigma_{S_{Q}}^2).
	\end{gathered}
\end{equation}

\noindent Note we have added the duration distribution parameters of each state to the list of parameters of the HMM. Specifically, the set of model parameters of the HSMM presented includes $\bm{\mu}, \bm{\sigma}^2,\bm\rho,\mathbf{P}$, $ \bm{\phi}$, $\mathbf{S}$, and $\bm{\tau}$.

\subsection{Use of covariates to model duration}
\label{subsec:cov}

Previous approaches have specified non-homogeneous HMM and HSMMs by modeling the parameters of the emission distribution or the probabilities of transition using covariates. 
We propose introducing non-homogeneity in the HSMM duration by letting the parameters of the state duration distribution vary in time as a function of covariates. If we let the duration distribution be a zero-truncated Poisson, we can define the duration parameter $\phi_{S_{q+1}}$ of the interval $q+1$ as a function of the covariate measurements observed prior to the transition at $T_{q}+1$. Notice that this specification enables the duration parameter to be both state-specific and vary in time. 

Let $\mathbf{X}$ be an $n \times r$ covariate matrix with rows corresponding to times $1$ to $n$, where $r$ is the number of covariates. Let $\bm\beta_{S_{q+1}}$ be an $(r+1)$-dimensional coefficient vector for state $S_{q+1}$ (accounting for an intercept in the model). Then the duration parameter for interval $q+1$, which we denote as $\phi_{S_{q+1}}(\mathbf{X}_{1:{T_{q}}},\bm \beta_{S_{q+1}})$, is a function of the covariate values observed up to time point $T_{q}$ (the first $T_q$ rows of $\textbf{X}$), and state specific coefficients $\bm{\beta}_{S_{q+1}}$. Here, $\phi_{S_{q+1}}(\mathbf{X}_{1:{T_{q}}},\bm \beta_{S_{q+1}})$ can take any functional form of the covariates as long as $\phi_{S_{q+1}} > 0$. For example, we can write the function as:

\begin{equation}
	\label{phif}
	\phi_{S_{q+1}}(\mathbf{X}_{1:{T_{q}}},\bm \beta_{S_{q+1}}) = g \left( \beta_{0,S_{q+1}} 
	+ \beta_{1,S_{q+1}}   f_1 \left( \mathbf x_{1,1:{T_{q}}} \right)   
	+ \cdots
	+ \beta_{r,S_{q+1}}   f_r \left( \mathbf x_{r,1:{T_{q}}} \right) 
	\right), 
\end{equation}

\noindent where $g(\cdot)$ is a specified function that ensures $\phi_{S_{q+1}} > 0$, and $f_1(\cdot), \dots, f_r(\cdot)$ can be any function of the covariates observed from time 1 to the time previous to the transition, $T_{q}$. The joint distribution, which now includes the state-specific duration parameter function, can be written as:

\begin{equation}
	\label{eqcov}
	\begin{gathered}
	  L(y_{\tau_1},\dots,y_{\tau_Q}, S_1,\dots,S_Q, \tau_1,\dots,\tau_Q \mid  \bm{\mu}, \bm{\sigma}^2,\bm\rho,\mathbf{P},\mathbf{B},\mathbf{X}) \hspace{40mm}\\
		= \rho_{S_1}  h_{S_1}\left( \tau_1 \mid \phi_{S_1}(\mathbf{X}_{0},\bm \beta_{S_1}) \right)  f(\mathbf{y}_{\tau_1}|\mu_{S_{1}},\sigma_{S_{1}}^2) \\
		\hspace{23mm}\times \hspace{2mm} \prod\limits_{q=2}^{Q}  h_{S_q}\left( \tau_q \mid \phi_{S_q}(\mathbf{X}_{1:{T_{q-1}}},\bm \beta_{S_q}) \right)   p_{S_{q-1},S_{q}}  f(\mathbf{y}_{\tau_q}|\mu_{S_{q}},\sigma_{S_{q}}^2),
	\end{gathered}
\end{equation}

\noindent where $\mathbf{X}_{0}$ are the initial values for the covariates, and $\mathbf{B}$ is the matrix of $\beta$-coefficients with M rows and the number of columns is the number of covariates, $r$, plus an intercept:

\begin{equation*}
	\mathbf{B} =
	\begin{pmatrix}
		\beta_{0,1} & \beta_{1,1} & \cdots & \beta_{r,1}\\
		\beta_{0,2} & \beta_{1,2} & \cdots & \beta_{r,2}\\
		\vdots  & \vdots  & \ddots & \vdots \\
		\beta_{0,M} & \beta_{1,M} & \cdots & \beta_{r,M}\\
	\end{pmatrix}.
\end{equation*}

\noindent That is, $\bm\beta^\prime_{S_q}$ is the row of $\mathbf{B}$ that corresponds to state $S_q$. For example, when the state in interval $q$ is 1, then $\bm\beta_{S_q=1} = \left( \beta_{0,1}, \beta_{1,1}, \dots , \beta_{r,1} \right)^\prime $.


\subsection{Estimation of model parameters}
\label{subsec:estim}
Model inference can be obtained in a Bayesian framework using Markov chain Monte Carlo (MCMC) and a Metropolis-within-Gibbs sampling algorithm (see Appendix \ref{sec_appsa} for the detailed sampling algorithm). To complete the model specification, we assign diffuse priors to the model parameters. The means of the emission distribution are assigned independent Normal priors,  $N\left(0, 10000 \right)$ and the variances are assigned inverse-Gamma priors, $IG\left(3,3 \right)$. The initial probabilities, as well as each of the rows of the transition matrix, have Dirichlet priors, $Dir(1,1,1)$ and $Dir(1,1)$, respectively. The coefficient parameters in the model for the state-specific durations are assumed to be independent and distributed as $N \left(0, 10000 \right)$.

The posterior distribution of the states, durations, and rest of the parameters of the HSMM can be summarized as:

\begin{equation}
	\begin{gathered}
		p(\bm{S},\bm{\tau},\bm{\mu},\bm{\sigma}^2,\bm{\rho},\mathbf{P},\mathbf{B} \mid \mathbf{y},\mathbf{X}) \propto 
		p(\mathbf{y} \mid \bm{S},\bm{\mu},\bm{\sigma}^2) \times p(\bm{\tau} \mid \bm{S},\mathbf{B},\mathbf{X}) \times p(\bm{S} \mid \bm{\rho},\mathbf{P}) \hspace{5mm} \\
		\times p(\bm{\mu} \mid \bm{\theta}_\mu,\bm{\lambda}_\mu^2)
		\times p(\bm{\sigma}^2 \mid \bm{\theta}_{\sigma^2},\bm{\lambda}_{\sigma^2})
		\times p(\mathbf{B} \mid \bm{\theta}_B, \bm{\lambda}_B^2)
		\times p(\bm{\rho}\mid \bm{\theta}_\rho) \times p(\bm{P} \mid \bm{\theta}_P).
	\end{gathered}
\end{equation}

\noindent The state means and variances, initial probabilities, and transition probabilities can be sampled from their full conditionals using a Gibbs update, whereas a Metropolis algorithm is needed for the duration distribution coefficients. 

\cite{economou_mcmc_2014} provide an MCMC implementation of the HSMM using a forward algorithm to estimate the parameters, which alleviates the need to sample the state sequence in the process. However, our model requires sampling the states in order to obtain inference on the parameters of the duration distributions. The state sequence in an HSMM can be sampled with the Gibbs sampler presented in \cite{johnson_bayesian_2012}, and we use it to sample the states in each iteration of our MCMC.
Additionally, the MCMC algorithm has to be adjusted to account for the conditional dependency in the observations. This modification is the inclusion of a subsampling approach that is discussed in the next section.

\subsection{Violation of the conditional independence assumption}
\label{subsec:sim}
We consider a simulation study to investigate the effect of the conditional independence assumption violation in the emission distribution parameter estimation. We simulate dependent data using an AR(1) model under several scenarios. Then, we fit an HSMM model to each generated dataset using a data subsampling approach, where at each iteration of the MCMC algorithm, an independent and random subset of the data is selected according to a specified sampling rate.

The different scenarios included in the simulation study are defined according to sample size, AR(1) autocorrelation parameter, and number of states. Three different number of states are considered ($M=$ 2, 3 and 4), while the possible values for the autocorrelation parameters are 0.25, 0.50, and 0.75, as well as a case with no autocorrelation. Although the number of simulated observations varies across all realizations, we consider two cases consisting of approximately 2500 and 5000 observations. Overall, 24 scenarios were considered, each with 100 realizations.

The procedure for simulating the data is as follows. First, the initial state, $S_1$, is sampled from its distribution. Second, the first duration, $\tau_1$, is generated from the corresponding zero-truncated Poisson distribution. The observations, $y_{1},y_{2},\cdots,y_{\tau_1}$, are generated from a Normal distribution with a specified autocorrelation. Then, the second state, $S_2$, is sampled conditional on $S_1$ according to the transition probability matrix $\textbf{P}$. After that, the duration is sampled as well as the observations emitted by that second state. The process continues until the specified sample size is reached.

Each of the simulated realizations is modeled with an HSMM assuming independence in the observations. For ease of computation, the states and other parameters are fixed and only the emission distribution parameters are estimated.

We investigate the benefit of the subsampling approach under various sample sizes, autocorrelation parameters, and number of states. In this approach, each iteration of the MCMC algorithm uses a different random subset of data according to a pre-specified percentage. The sampling rates to be considered are 100, 90, 80, $\dots$, 10\%. The 90\% credible intervals (CI) for the mean and variance parameters for all 100 datasets across all scenarios and sampling rates are then calculated. We assess the subsampling approach by comparing the empirical coverage, and determine the preferred data sampling rate as the one that results in the nominal coverage.

The results of the simulation study are provided in Appendix \ref{sec_appperc}. Overall, the correlation in the data affects the estimation of the emission distribution parameters, but the effect can be reduced by subsampling during the model fitting procedure. We use this result for the phycocyanin example in Section \ref{sec:appl}. 

\section{Application to Lake Mendota phycocyanin data}
\label{sec:appl}
The HSMM specified in Equation \ref{eqcov} was used to model the hourly maximum standardized levels of phycocyanin in Lake Mendota, for the period April 11 to November 15, 2018. The duration (hours) in each state is modeled using a zero-truncated Poisson distribution, where the duration parameter is defined as a function of four covariates. These include temperature, wind speed, relative humidity, and PAR. 
Including the intercept term, this results in five coefficient parameters for each state. Following the notation in Equation \ref{phif}, the duration parameter function in this application is defined as:

\begin{equation} 
	\label{phifapp} 
	\begin{gathered} 
		\phi_{S_{q+1}}(\mathbf{X}_{1:T_{q}},\bm \beta_{S_{q+1}}) = \exp \left [ \beta_{0,S_{q+1}} + \beta_{1,S_{q+1}} x_{1,T_q} + \beta_{2,S_{q+1}} x_{2,T_q} + \beta_{3,S_{q+1}} x_{3, T_q}  + \beta_{4,S_{q+1}} x_{4, T_q} \right] . 
	\end{gathered} 
\end{equation}

To determine the optimal subsampling rate for our analysis, we first ran the algorithm without subsampling and determined an approximate value for the size of the segments. The durations varied by state and through time, but the average duration ranged from 21 to 55. We divided the data into segments to resemble the groups of emitted data by a state and calculated the mean autocorrelation across all groups. Considering group sizes of 21 to 55 resulted in autocorrelation values that ranged from 0.653 to 0.797, which we then compared to values in Table \ref{table_coveragesmu3s}. 
Using $\psi=0.75$ suggests a subsampling rate of approximately 30\% for estimating the emission distribution parameters.

The number of states in HMMs and HSMMs has to be chosen {\it a priori} and is usually determined through information criteria or expert knowledge \citep{liu_bayesian_2020}. However, the information criteria can lead to selecting a higher number of states given that they under-penalize model complexity, as mentioned in \cite{celeux_selecting_2008} and in the discussion of \cite{spiegelhalter_bayesian_2002}. In our case, the deviance information criterion (DIC) always selected the model with the largest number of states, motivating the need to consider other measures for model selection. Specifically, we considered a convergence diagnostic as an alternative for choosing the appropriate number of states. We obtained multiple independent parameter chains for cases with 2, 3, 4, and 5 states and assessed convergence with the potential scale reduction factor (PSRF) and its multivariate equivalent (MPSRF) discussed in \cite{brooks_general_1998}. The MPSRF for the 2, 3, 4, and 5 state models were 1.83, 1.08, 1.19, and 6.84, respectively. In addition, in the 3 state case the upper bound of the PSRF value for each model parameter was less than 1.2. Thus, we selected the 3 state model as our model for further analysis.

The MCMC algorithm was run for 24000 iterations. The first 4000 iterations were obtained using an adaptive random walk Metropolis algorithm for the duration distribution coefficients. These iterations were used to select the proposal variances for the random walk and then discarded. The remaining 20000 iterations were obtained based on these fixed proposal variances and these samples were used for parameter inference.

\begin{figure}[H]
	\begin{center}
		\includegraphics[scale=.5]{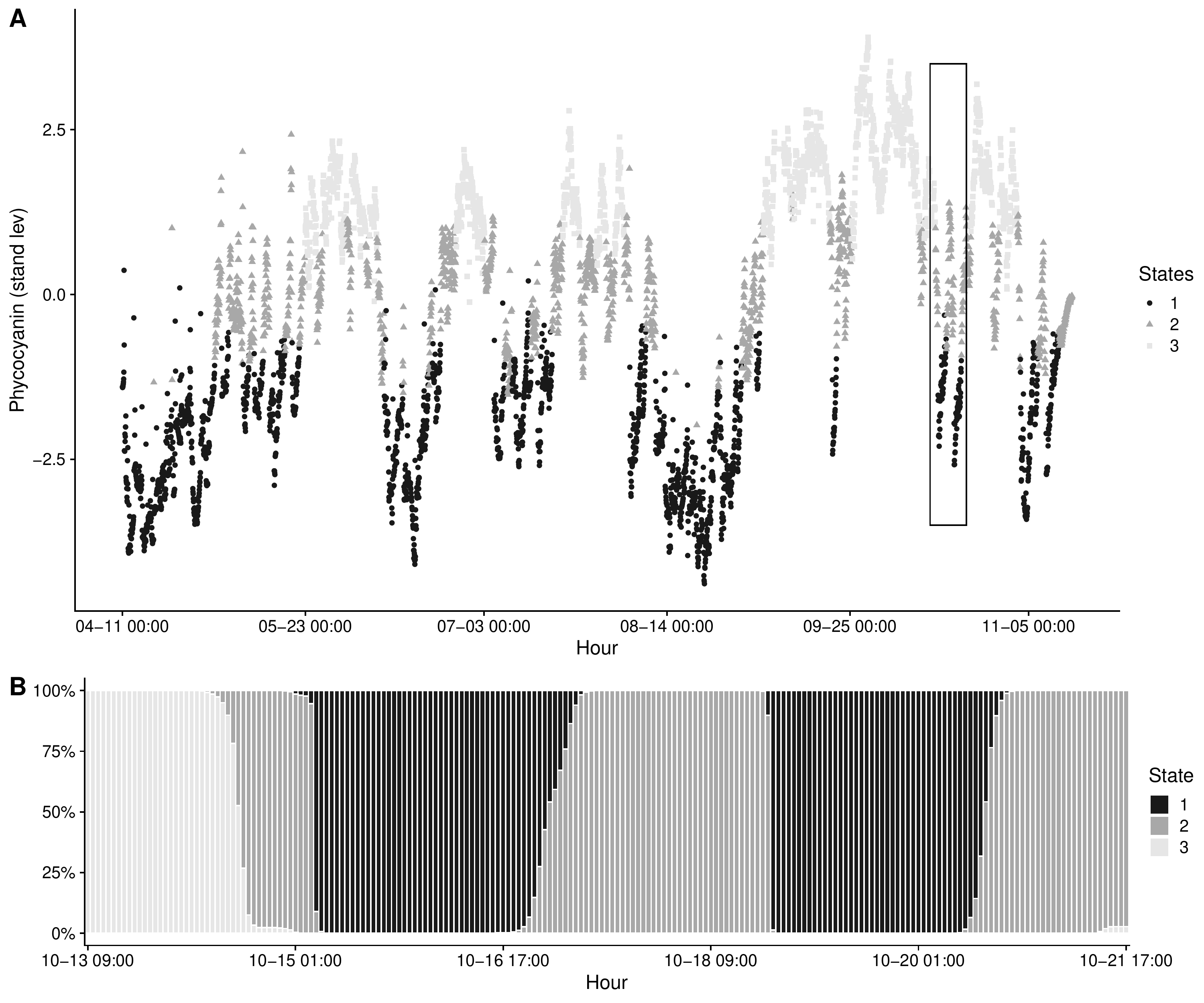}
		\caption{Phycocyanin standardized levels classified by latent states of cyanobacteria. Panel \textbf{A}. Full state sequence. For each time point, the state is the mode obtained among all iterations. Panel \textbf{B} zooms in the period enclosed in the rectangle above. A percent stacked bar for each time point shows the relative distribution of the states sampled in the iterations. \label{Fig:G3states}}
	\end{center}
\end{figure}

The posterior mean and 95\% credible intervals for the emission distribution parameters in each of the states is presented in Table \ref{table_meanG3}. There is a clear distinction between the mean phycocyanin in each state since none of the credible intervals overlap. The three states represent low, medium, and high cyanobacteria states. The variability in the middle state is notably higher, and the wide range of this state can be seen in Figure \ref{Fig:G3states}. These low, medium and high states of cyanobacteria can be associated with the regimes found in \cite{carpenter_stochastic_2020}. Recall that they identified two stable states, which are comparable to our low and high states (S1 and S3), while the shifts in between these two regimes correspond to our middle state (S2).

\begin{table}[H]
	\centering
	\caption{Posterior mean and 95\% CI of the emission distribution parameters}
	\label{table_meanG3}
	\begin{tabular}{lccc}
		\hline
		State & Mean & Variance \\
		\hline
		S1 & -2.18 (-2.30, -2.02) & 0.76 (0.64, 0.95)  \\ 
		S2 & 0.09 (-0.06, 0.31) & 0.45 (0.35, 0.56) \\ 
		S3 & 1.75 (1.65, 1.86) & 0.45 (0.38, 0.54) \\ 
		\hline
	\end{tabular}
\end{table}

\bigskip

The posterior probabilities of transitioning are shown in Table \ref{table_probG3}. Overall, it is more likely there is a transition between adjacent states (e.g., 1 to 2) than a jump from 1 to 3. This is also true for transitions from higher to lower cyanobacteria states (e.g., 3 to 2). The transition between adjacent states is expected given that the middle state represents a passing state from regimes of cyanobacteria concentration.

\bigskip

\begin{table}[H]
	\centering
	\caption{Posterior mean and 95\% CI for the state transition probabilities.}
	\label{table_probG3}
	\begin{tabular}{lcccc}
		\hline
		Transition & State 1 & State 2 & State 3  \\ 
		\hline
		State 1 & -- 				& 0.96 (0.86, 1) 	 & 0.04 (0, 0.14)\\ 
		State 2 & 0.59 (0.43, 0.74) & --				 & 0.41 (0.26, 0.57)\\ 
		State 3 & 0.09 (0, 0.28) & 0.91 (0.72, 1)  & -- \\ 
		\hline
	\end{tabular}
\end{table}


Table \ref{table_durationratesG3} provides the number of segments in each state as well as summaries of the duration parameters and durations for each state. 
The second state has more segments, yet the average duration parameter for this state is much smaller than the other two states.
The first and third state have fewer segments, but both the mean and variation in the duration parameter is greater than for the second state. 
The variation in the duration parameters both within and between states signifies the importance of modeling the durations using covariates and state-specific parameters.

\begin{table}[H]
	\centering
	\caption{Segments and duration parameter statistics}
	\label{table_durationratesG3}
	\begin{tabular}{cccccccc}
		\hline
		{\multirow{2}{*}{State}} &
		{\multirow{2}{*}{\# segments}} &
		\multicolumn{3}{c}{Duration parameter}&
		\multicolumn{3}{c}{Duration (hours)}\\ 	
		\cline{3-8}
		& & Mean & Minimum & Maximum &Mean & Minimum & Maximum \\
		\hline
		S1 & 26  & 84.2 	& 33.7	& 198.2  	& 84 	& 17 & 234 \\
		S2 & 42  & 33.2 	& 14.7  & 65.9  	& 33  & 1  & 93\\ 
		S3 & 17  & 102.4 	& 69.0 	& 139.3 	& 102.4 & 42 & 177\\ 
		\hline
	\end{tabular}
\end{table}

The posterior means and credible intervals for the coefficients of the state-specific duration distribution parameters are given in Table \ref{table_betaG3}, with the significant coefficients presented in bold font. The coefficients of the duration distribution provide information about the average hourly duration in the states. 

Air temperature is significant in capturing the variation in the duration in the low cyanobacteria state and is inversely related to duration. That is, when the temperature before a transition to the low state is warm, the duration in that state is shorter. 
Wind speed is also significant in the low cyanobacteria state. When wind speeds prior to a transition to the low cyanobacteria state are high, we anticipate a shorter duration in that state. Relative humidity is a significant predictor of duration in the second state. Lastly, the photosyntetically-active radiation covariate is related to the first and second state. The positive coefficient for the first state indicates that when PAR is at higher levels before there is a transition to the lower cyanobacteria state, we expect an increase in the duration. The relation is inverse to the duration in the intermediate state of cyanobacteria concentration.

\begin{table}[H]
	\centering
	\caption{Posterior mean and 95\% CI of the duration parameter coefficients.}
	\label{table_betaG3}
	\begin{tabular}{lccc}
		\hline
		Variable & State 1 & State 2 & State 3\\ 
		\hline
		Intercept 			& \bf 4.37 (4.22, 4.57) 	& \bf 3.43 (3.22, 3.64) 	& \bf 4.57 (4.26, 4.81) \\ 
		Temperature 		& \bf -0.16 (-0.28, -0.02) 	& 0.03 (-0.12, 0.20) 		& 0.08 (-0.18, 0.30) \\ 
		Wind speed 			& \bf -0.19 (-0.34, -0.03) 	& 0.07 (-0.08, 0.21) 		& 0.04 (-0.15, 0.20) \\ 
		Relative humidity 	& 0.04 (-0.21, 0.19) 		& \bf -0.37 (-0.57, -0.18) 	& 0.11 (-0.08, 0.27) \\ 
		PAR 				& \bf 0.32 (0.16, 0.48) 	& \bf -0.28 (-0.51, -0.07) 	& -0.02 (-0.29, 0.23) \\
		\hline
	\end{tabular}
\end{table}



\section{Discussion}
\label{sec:disc}
An extension of the HSMM was presented and applied to a high-frequency environmental dataset. We introduced non-homogeneity into the duration distribution of the HSMM using time-varying covariates.
In our example, the model enabled the characterization of cyanobacteria concentration in a lake. The three states for phycocyanin represent low, medium, and high levels, and their mean and variance were estimated.

Using a zero-truncated Poisson distribution for the duration (hours) in each state, we investigated the variation in time spent in each state as a function of time-varying covariates. 
Important differences were detected in the relationship between the duration of time in the states and the covariates.  
The variability of the duration parameters in the different segments supports the introduction of non-homogeneity in the HSMM for this application.

We extended the results obtained in \cite{carpenter_stochastic_2020} by modeling the duration in the regimes of cyanobacteria concentration. With a different modeling approach, we identified similar states of cyanobacteria levels, and determined there are weather covariates associated with the duration in each of those states. Although not demonstrated here, this model could be applied to obtain predictions of the next states in the sequence and their expected duration.

The inference obtained from the duration in this model can be potentially associated with ecological resilience. \cite{arani_exit_2021} propose to measure resilience, the maximum perturbation that a system endures without transitioning to another state, with life expectancy. They present how to fit a Langevin equation to time series data to capture the different forces that affect a system, and obtain the mean exit time from a state to quantify life expectancy. The information provided by our model in terms of duration in a particular state before transitioning to another can be explored to measure life expectancy as well. 

Other approaches exist for introducing non-homogeneity into HSMMs. For example, parameters corresponding to transition probabilities could also be modeled as a function of covariates.
However, the focus of the analysis presented here was in capturing the variation in the duration of time in each state. 
Given that a direct transition from the low to the high state, or vice versa, is unlikely, it was not necessary to model the transition probabilities in terms of covariates for this application.

The subsampling approach used in the estimation of the means and variances helped reduce the effect of dependence in the observation sequence. 
Our simulation study provided guidance as to the level of subsampling necessary to properly account for uncertainty in the model parameters. 
An indirect result of the subsampling was reduced computation time. 
Since high-frequency data are becoming increasingly common, additional research will focus on accounting for conditional dependence in the data.


\section*{Acknowledgements}

This research was funded by the Macrosystems Biology Program in the Emerging Frontiers Division of the Biological Sciences Directorate at the U.S. National Science Foundation (EF-1638554 and EF-1638550). Additional support was provided by NSF Award DMS-1811745 and by the Office of International Affairs and External Cooperation, University of Costa Rica.

\bibliographystyle{apalike}
\bibliography{phyco_new}

\begin{appendix}

\newpage
\newgeometry{left=1in, right=1in,top=1in,bottom=1in}	
	\counterwithin{table}{section}	
	
	\section{Sampling algorithm}
	\label{sec_appsa}
	
	\begin{algorithm}
		\caption{MCMC sampling algorithm}
		\bigskip
		\textbf{Initial values}
		\begin{algorithmic}[1]
			\State Define initial values for parameters $\bm{\rho}^{(0)}, \textbf{P}^{(0)}, \textbf{B}^{(0)}$.
			\State Generate the state sequence $\textbf{S}^{(0)}$  as follows:
			\begin{algsubstates}
				\smallskip
				\State Sample the first state $S_{q=1}^{(0)}$ using $\bm{\rho}^{(0)}$.
				\smallskip
				\State Calculate $\phi_{S_{q=1}}^{(0)}$ using $\textbf{X}$ and $\bm{\beta}_{S_{q=1}}^{(0)}$.
				\smallskip
				\State Sample $\tau_1^{(0)}$ from a zero-truncated Poisson with parameter $\phi_{S_{q=1}}^{(0)}$.
				\medskip
				\State Define $\textbf{S}_{1:T_1}=S_{q=1}$.
				\smallskip
				\State Sample $S_{q=2}^{(0)}$ conditional on $S_{q=1}^{(0)}$ using $\textbf{P}^{(0)}$.
				\smallskip
				\State Calculate $\phi_{S_{q=2}}^{(0)}$ using $\textbf{X}$ and $\bm{\beta}_{S_{q=2}}^{(0)}$.
				\smallskip
				\State Sample $\tau_2^{(0)}$ from a zero-truncated Poisson with parameter $\phi_{S_{q=2}}^{(0)}$.
				\medskip
				\State Define $\textbf{S}_{{T_1}+1:T_2}=S_{q=2}$.
				\medskip
				\State Continue until $T_q=n$.
			\end{algsubstates}
			\State Calculate $\bm{\mu}^{(0)}$ and ${\bm{\sigma}^2}^{(0)}$ based on $\textbf{S}^{(0)}$.
		\end{algorithmic}
		\bigskip
		
		\textbf{Iterations}
		\begin{algorithmic}[1]
			\For{iteration $l=1,2,\dots$}
			
			\State Update $\bm{\rho}^{(l-1)}$ using Gibbs sampling:
			\begin{equation*}
				\bm{\rho}^{(l)} \sim Dir\left( I\left(S_{1}^{(l-1)}=1 \right)  + \theta_{\rho_1}, \dots, I\left(S_{1}^{(l-1)}=M \right)  + \theta_{\rho_M} \right) ,
			\end{equation*}
			where $I(\cdot)$ is the indicator function.	
			
			\State Update the $j$-th row of $\textbf{P}^{(l-1)}$ for $j=1, 2, \dots, M$, using Gibbs sampling:
			\begin{equation*}
				\textbf{P}_j^{(l)} \sim Dir\left( n_{j1}  + \theta_{P_{j1}}, \dots, n_{jM}  + \theta_{P_{jM}} \right) ,
			\end{equation*}
			where $n_{jk}= \sum\limits_{q=1}^{Q-1} I\left(S_{q}^{(l-1)}=j,S_{q+1}^{(l-1)}=k \right) $ is the total number of transitions from state $j$ to state $k$.

			\algstore{myalg}
		\end{algorithmic}
	\end{algorithm}
	
	\clearpage
	
	\begin{algorithm}
		\ContinuedFloat
		\caption{MCMC sampling algorithm (continued)}
		\begin{algorithmic}
			\algrestore{myalg}
			
			\State Update $\bm{\beta}_j^{(l-1)}$ for $j=1, 2, \dots, M$ using random-walk Metropolis. Sample $\textbf{z}\sim N(\textbf{0},\kappa_{\beta_j}^2 \textbf{I}_{r+1})$, where $ I_{r+1}$ is an identity matrix of order $r+1$, and define the proposal vector $\bm{\beta}_j^{(*)}=\bm{\beta}_j^{(l-1)} + \textbf{z}$. Then calculate the Metropolis ratio as:
			\begin{equation*}
				m_{\bm{\beta}_j}= \left( \frac{ \prod \limits_{q=1}^Q ZTP \left(\tau_{q}^{(l-1)} \mid \phi_q^{(*)} \right)}{\prod \limits_{q=1}^Q ZTP \left(\tau_{q}^{(l-1)} \mid \phi_q^{(l-1)} \right)} \right) \times \left( 	\frac{N\left(\bm{\beta}_j^{(*)} \mid \bm{\theta}_{\beta_j}, \lambda_{\beta_j}^2 \textbf{I}_{r+1} \right) }{N\left(\bm{\beta}_j^{(l-1)} \mid \bm{\theta}_{\beta_j}, \lambda_{\beta_j}^2 \textbf{I}_{r+1} \right) } \right)
				,	
			\end{equation*}
			and if $u < m_{\bm{\beta}_j}$, with $u\sim Unif(0,1)$, let  $\bm{\beta}_j^{(l)}=\bm{\beta}_j^{(*)}$, and update $\bm\phi^{(l-1)}$, $\bm\phi^{(l)}=\bm\phi^{(*)}$.
			
			\State Update $\textbf{S}^{(l-1)}$ and $\bm\tau^{(l-1)}$ with the sampler in \cite{johnson_bayesian_2012} for the finite HSMM. 
			
			\State Subsample the observations. Randomly sample the observations $\textbf{y}$ according to the specified sampling rate, and use the new observation vector $\tilde{\textbf{y}}$ of size $\tilde{n}$ in steps 10 and 11.	
			
			\State Update $\mu_j^{(l-1)}$ for $j=1, 2, \dots, M$ using Gibbs sampling and the subsampling approach: 
			\begin{equation*}
				\mu_j \sim {N} \left(  \frac{\frac{\sum \limits_{i=1}^{n} y_i  {I({S}_i=j, y_i \in \tilde{\textbf{y}})}}{{\sigma_{j}^2}^{(l-1)}} + \frac{\theta_\mu}{\lambda_\mu^2}}{ \frac{\tilde{n}_{j}}{{\sigma_{j}^2}^{(l-1)}} + \frac{1}{\lambda_\mu^2}}, \frac{1}{ \frac{\tilde{n}_{j}}{{\sigma_{j}^2}^{(l-1)}} + \frac{1}{\lambda_\mu^2}} \right),
			\end{equation*}
			where $I(\cdot)$ is the indicator function and $\tilde{n}_j$ is the total number of observations of vector $\tilde{\textbf{y}}$ emitted by state $j$.
			
			\State Update ${\sigma_j^2}^{(l-1)}$ for $j=1, 2, \dots, M$  using Gibbs sampling and the subsampling approach:
			\begin{equation*}
				\sigma_{j}^2 \sim IG \left( \theta_{\sigma^2} + \frac{\tilde{n}_j}{2} ,   \lambda_{\sigma^2} + \frac{1}{2}\sum \limits_{i=1}^{n} \left( y_i  {I({S}_i=j, y_i \in \tilde{\textbf{y}})} - \mu_j^{(l-1)} \right) ^2   \right) .
			\end{equation*}
			
			\State Save $\bm{\rho}^{(l)}, \textbf{P}^{(l)}, \textbf{B}^{(l)}, \bm\phi^{(l)},\textbf{S}^{(l)},\bm\tau^{(l)}, \bm{\mu}^{(l)}$ and ${\bm{\sigma}^2}^{(l)}$.
			
			\EndFor
		\end{algorithmic}
	\end{algorithm}
	
\restoregeometry

\section{Simulation study results}
\label{sec_appperc}

The simulation study was developed for 2, 3 and 4 states. The empirical coverage is calculated as the percentage of 90\% CIs that captured the true parameter values. Tables \ref{table_coveragesmu2s} to \ref{table_coveragesmu4s} present the empirical coverage for the different autocorrelation parameters, sample size and sampling rates utilized. This coverage corresponds to the average coverage of the different state means. The empirical coverage is similar for the two sample size and the number of states scenarios. 

There are two important results provided in the tables. First, the more correlated the data are, the worse we do in recovering the true parameter values, as indicated by the first row of the tables where no subsampling was used. Second, the empirical coverage increases as the sampling rate decreases. As we reduce the percent of data used, we are able to reduce the dependence in the data and thus improve our estimates of uncertainty. For a case where the autocorrelation is approximately 0.75 and the whole dataset is utilized (sampling rate = 100\%), then the coverage is low, indicating the need to use a smaller sampling rate. With data having autocorrelation close to 0.25, we obtain nominal coverage of the emission distribution means when using approximately 80\% of the data points in the MCMC iterations.

\begin{table}[H]
	\centering
	\caption{Coverage percentage of the emission distribution means, 2 states case.}
	\label{table_coveragesmu2s}
	\begin{tabular}{cccccccc}
		\hline 
		{\multirow{2}{*}{Sampling rate \%  }} &
		\multicolumn{3}{c}{$\psi$ (n$\approx$2500)}& &
		\multicolumn{3}{c}{$\psi$ (n$\approx$5000)}\\
		\cline{2-4} 
		\cline{6-8} 	
		& 0.25 & 0.50 & 0.75 & & 0.25 & 0.50 & 0.75\\
		\hline
		100 & 78 & 68 & 44 & & 76 & 68 & 55 \\ 
		90 & 83 & 73 & 50 & & 81 & 74 & 60 \\ 
		80 & 87 & 80 & 54 & & 86 & 76 & 66 \\ 
		70 & 92 & 84 & 62 & & 90 & 81 & 72 \\ 
		60 & 96 & 85 & 67 & & 93 & 84 & 77 \\ 
		50 & 98 & 88 & 70 & & 94 & 88 & 81 \\ 
		40 & 98 & 93 & 76 & & 97 & 94 & 86 \\ 
		30 & 100 & 96 & 88 & & 100 & 98 & 91 \\ 
		20 & 100 & 99 & 92 & & 100 & 100 & 96 \\ 
		10 & 100 & 100 & 100 & & 100 & 100 & 100 \\ 
		\hline
	\end{tabular}
\end{table}

\begin{table}[H]
	\centering
	\caption{Coverage percentage of the emission distribution means, 3 states case.}
	\label{table_coveragesmu3s}
	\begin{tabular}{cccccccc}
		\hline 
		{\multirow{2}{*}{Sampling rate \%  }} &
		\multicolumn{3}{c}{$\psi$ (n$\approx$2500)}& &
		\multicolumn{3}{c}{$\psi$ (n$\approx$5000)}\\
		\cline{2-4} 
		\cline{6-8} 	
		& 0.25 & 0.50 & 0.75 & & 0.25 & 0.50 & 0.75\\
		\hline
		100 & 77 & 68 & 50 & & 82 & 67 & 50 \\
		90 & 82 & 75 & 56 & & 86 & 73 & 55 \\
		80 & 89 & 80 & 60 & & 90 & 76 & 60 \\
		70 & 92 & 84 & 65 & & 92 & 81 & 64 \\
		60 & 96 & 87 & 71 & & 94 & 86 & 72 \\
		50 & 99 & 92 & 79 & & 96 & 92 & 79 \\
		40 & 99 & 96 & 84 & & 99 & 96 & 84 \\
		30 & 100 & 99 & 90 & & 100 & 98 & 90 \\
		20 & 100 & 100 & 95 & & 100 & 100 & 95 \\
		10 & 100 & 100 & 100 & & 100 & 100 & 99 \\
		\hline
	\end{tabular}
\end{table}

\begin{table}[H]
	\centering
	\caption{Coverage percentage of the emission distribution means, 4 states case.}
	\label{table_coveragesmu4s}
	\begin{tabular}{cccccccc}
		\hline 
		{\multirow{2}{*}{Sampling rate \%  }} &
		\multicolumn{3}{c}{$\psi$ (n$\approx$2500)}& &
		\multicolumn{3}{c}{$\psi$ (n$\approx$5000)}\\
		\cline{2-4} 
		\cline{6-8} 	
		& 0.25 & 0.50 & 0.75 & & 0.25 & 0.50 & 0.75\\
		\hline
		100 & 80 & 66 & 52 & & 80 & 67 & 53 \\ 
		90 & 86 & 70 & 57 & & 86 & 72 & 56 \\ 
		80 & 89 & 75 & 61 & & 89 & 78 & 58 \\ 
		70 & 94 & 81 & 66 & & 93 & 83 & 62 \\ 
		60 & 96 & 85 & 69 & & 95 & 86 & 69 \\ 
		50 & 98 & 90 & 75 & & 97 & 90 & 76 \\ 
		40 & 100 & 96 & 82 & & 98 & 95 & 83 \\ 
		30 & 100 & 99 & 87 & & 99 & 98 & 90 \\ 
		20 & 100 & 100 & 94 & & 100 & 100 & 96 \\ 
		10 & 100 & 100 & 99 & & 100 & 100 & 99 \\ 
		\hline
	\end{tabular}
\end{table}

\end{appendix}

\end{document}